\documentclass[a4paper,12pt,eqno]{article}
\usepackage{amssymb}

\begin{document}

\title{Relation between coined quantum walks and quantum cellular automata}

\author{\\
\\
Masatoshi Hamada, Norio Konno, and Etsuo Segawa, \\
Department of Applied Mathematics, \\
Yokohama National University \\}
\date{}
\maketitle

\newcommand{\ket}[1]{|#1\rangle}
\newcommand{\bra}[1]{\langle#1|}
\newcommand{\U}{\bar{U}}
\newcommand{\braa}{\langle}
\newcommand{\kett}{\rangle}

\begin{small}
\noindent
{\bf Abstract}. Motivated by the recent work of Patel et al., this paper clarifies a connection between coined quantum walks and quantum cellular automata in a general setting. As a consequence, their result is naturally derived from the connection. 
\end{small}

\section{Introduction}
Very recently Patel {\it et al.} \cite{Patel} constructed a quantum walk (QW) on a line without using a coin toss instruction, and analyzed the asymptotic behavior of the walk on the line and its escape probability with an absorbing wall. In fact the QW investigated by them can be considered as a class of quantum cellular automata on the line (see \cite{MeyerA,MeyerB,MeyerC}, for examples), so we call their non-coined QW a {\it quantum cellular automaton} (QCA) in this paper. On the other hand, a usual QW with a coin toss instruction was introduced and intensively studied by Ambainis {\it et al.} \cite{Ambainisetal}, (see Refs. \cite{Kempe,Tregenna,Ambainis} for reviews of the QW). Here we call the QW a {\it coined QW.} At a first glance, the QCA looks different from the coined QW. However, we show that there exists a one-to-one correspondence between them in a more general setting. The purpose of the present paper is to clarify this connection. Once the connection is well understood, the result by Patel {\it et al.} \cite{Patel} is straightforwardly obtained.

The rest of this paper is organized as follows. Section 2 is devoted to the definition of the QCA. In Section 3, the definitions of A-type and B-type QWs are presented. In Section 4 (resp. 5), we describe a connection between QCAs and A-type (resp. B-type) QWs. Section 6 gives a relation between QCAs and two-step coined QWs. Finally, we discuss the case given by Patel {\it et al.} \cite{Patel}.

\section{Definition of QCA}
Let ${\bf Z}$ (resp. ${\bf Z_+}$) denote the set of (resp. non-negative) integers and ${\bf C}$ indicate the collection of complex numbers. We define the dynamics of a one-dimensional QCA including the model investigated by Patel {\it et al.} \cite{Patel}. (As for recent reviews on QCAs, see Aoun and Tarifi \cite{Aoun}, Schumacher and Werner \cite{Schumacher}.) Let $\eta^{(m)} _k (n) (\in {\bf C})$ be the amplitude of the QCA at time $n \in {\bf Z_+}$ and at location $k \in {\bf Z}$ starting from $m \in {\bf Z}$, that is, $\eta^{(m)} _m (0)=1$ and $\eta^{(m)} _k (0) =0$ if $k \not=m$. Moreover, let 
\begin{eqnarray*}
x_k ^{(m:\pm)} (n) = | \alpha \eta^{(m)} _k (n) + \beta \eta^{(m \pm 1)} _k (n) |^2 
\label{eqn:norio}
\end{eqnarray*}
and
\begin{eqnarray*}
x^{(m:\pm)} (n) =  ( x_k ^{(m:\pm)} (n) : k \in {\bf Z})
\label{eqn:noriohe}
\end{eqnarray*}
where $\alpha, \beta \in {\bf C}$ with $|\alpha|^2 + |\beta|^2 = 1.$ As we will show later, the $x_k ^{(m:\pm)} (n)$ is equivalent to a probability distribution of a coined QW at time $n$, where a pair $(\alpha,\beta)$ corresponds to an initial qubit state of the QW. The evolution of the QCA on the line is given by 
\begin{eqnarray*}
\eta^{(m)} (n+1) = \overline{U} \eta^{(m)} (n)
\end{eqnarray*}
where $\overline{U}$ is the unitary matrix 
\begin{eqnarray*}
\overline{U} \> = \> 
\bordermatrix{
   & \ldots & -3 & -2 & -1 & \>\>\>0 & +1 & +2 & +3 & +4 &\ldots \cr
\vdots & \ddots & \cdot  &  \cdot &  \cdot &  \cdot & \cdot &  \cdot &  \cdot & \cdot & \ldots \cr
-3 & \ldots &  b &  a &  0 & 0 &  0 &  0 &  0 & 0 & \ldots \cr
-2 & \ldots &  a &  b &  c & d &  0 &  0 &  0 & 0 & \ldots \cr
-1 & \ldots &  d &  c &  b & a &  0 &  0 &  0 & 0 & \ldots \cr
\>\>\> 0  &  \ldots & 0 &  0 &  a & b & c & d  & 0 & 0 & \ldots \cr
+1  &  \ldots & 0 &  0 &  d & c & b & a & 0 & 0 & \ldots \cr
+2  &  \ldots & 0 &  0 &  0 & 0 & a & b & c & d & \ldots \cr
+3  &  \ldots & 0 &  0 &  0 & 0 & d & c & b & a & \ldots \cr
+4  &  \ldots & 0 &  0 &  0 & 0 & 0 & 0 & a & b & \ldots \cr
\vdots & \ldots & \cdot  &  \cdot &  \cdot &  \cdot & \cdot &  \cdot &  \cdot & \cdot & \ddots 
}
\label{eqn:masae}
\end{eqnarray*}
with $a,b,c,d \in {\bf C}$ and $\eta^{(m)} (n)$ is the configuration
\begin{eqnarray*}
\eta^{(m)} (n) = {}^t 
(\ldots, \eta_{-1} ^{(m)} (n), \eta_{0} ^{(m)} (n), \eta_{+1} ^{(m)} (n), \ldots)
\end{eqnarray*}
for any $n \in {\bf Z}_+,$ here $t$ indicates the transposed operator. Let $|| u || = \sum_{k= - \infty} ^{\infty} | u _k |^2.$ The unitarity of $\overline{U}$ ensures that if $|| \eta^{(m)} (0) ||=1,$ then $|| \eta^{(m)} (n)||=1$ for any $n \in {\bf Z}_+.$ Furthermore, if $|| x^{(m:\pm)} (0) ||=1,$ then $|| x^{(m:\pm)} (n)||=1$ for any $n \in {\bf Z}_+.$ A little algebra reveals that $\overline{U}$ is unitary if and only if 
\begin{eqnarray}
&& |a|^2 + |b|^2  + |c|^2 + |d|^2 = 1, 
\label{eqn:yuki}
\\
&& a \overline{d} + \overline{a} d + b \overline{c} + \overline{b} c = 0, 
\label{eqn:aki}
\\
&& a \overline{c} + b \overline{d} = 0, 
\label{eqn:naomi}
\\
&& a \overline{b} + \overline{a} b = 0, 
\label{eqn:kazue}
\\
&& c \overline{d} + \overline{c} d = 0
\label{eqn:masako}
\end{eqnarray}
where $\overline{z}$ is a complex conjugate of $z \in {\bf C}.$ Here we consider $a,b,c,d$ satisfying Eqs. (\ref{eqn:yuki}) - (\ref{eqn:masako}). Trivial cases are ``$|a|=1, b=c=d=0$'', ``$|b|=1, a=c=d=0$'', ``$|c|=1, a=b=d=0$'', and ``$|d|=1, a=b=c=0$". For other cases, we have the following five types:
\par\noindent
Type I: $|b|^2+|c|^2=1, \> b \overline{c} + \overline{b} c = 0, \> bc \not=0, 
\> a=d=0.$
\par\noindent
Type II: $|a|^2+|b|^2=1, \> a \overline{b} + \overline{a} b = 0, \> ab \not=0, \> c=d=0.
$\par\noindent
Type III: $|c|^2+|d|^2=1, \> c \overline{d} + \overline{c} d = 0, \> cd \not=0, \> a=b=0.$
\par\noindent
Type IV: $|a|^2+|d|^2=1, \> a \overline{d} + \overline{a} d = 0, \> ad \not=0, \> b=c=0.$
\par\noindent
Type V: $a,b,c,d$ satisfying Eqs. (\ref{eqn:yuki}) - (\ref{eqn:masako}) and $abcd \not=0.$
\par
Let $\hbox{supp}[x^{(m:\pm)} (n)] = \{ k \in {\bf Z} : x^{(m:\pm)} _k (n) > 0 \}$. Then, it is easily seen that for any $n \in {\bf Z}_+$, $\hbox{supp}[x^{(0:\pm)} (n)] \subset \{-2,-1,0,1\}$ in Type I, and $\hbox{supp}[x^{(0:\pm)} (n)] \subset \{-1,0,1,2\}$ in Type II. So both Types I and II are also trivial cases. To investigate non-trivial Types III - V, we introduce a coined QW in the next section. 

We see that a direct computation implies that $(a,b,c,d)$ satisfying Eqs. (\ref{eqn:yuki}) - (\ref{eqn:masako}) has the following form:
\par\noindent
\begin{eqnarray}
(a,b,c,d) = 
e^{i \delta} (\cos \theta \cos \phi, -i \cos \theta \sin \phi, 
 \sin \theta \sin \phi, i \sin \theta \cos \phi)
\label{eqn:nobunaga}
\end{eqnarray}
where $\theta, \phi, \delta \in [0, 2 \pi).$ From now on, we assume that $(a,b,c,d)$ has the above form. Remark that the case studied by Patel {\it et al.} is $\delta = \pi/2$ and $\theta = \phi = \pi/4,$ that is, $(a,b,c,d)=(i/2,1/2,i/2,-1/2),$ which belongs to Type V.

\section{Definition of coined QW}
The time evolution of one-dimensional both A-type and B-type coined QWs is given by the following unitary matrix:
\begin{eqnarray*}
U=
\left(
\begin{array}{cc}
a' & b' \\
c' & d'
\end{array}
\right)
\end{eqnarray*}
\par\noindent
where $a',b',c',d' \in {\bf C}$. So we have $|a'|^2 + |b'|^2 =|c'|^2 + |d'|^2 =1, \> a' \overline{c'} + b' \overline{d'}=0, \> c'= - \triangle \overline{b'}, \> d' = \triangle \overline{a'},$ where $\triangle = \det U = a'd' - b'c'$ with $|\triangle|=1.$
\par
Let $\ket{L} = {}^t(1,0)$ and $\ket{R} = {}^t(0,1).$ For an A-type coined QW, each coin performs the evolution:
\begin{eqnarray*}
&& \ket{L} \quad \to \quad U \ket{L}=a' \ket{L} + c' \ket{R}, \\
&& \ket{R} \quad \to \quad U \ket{R}=b' \ket{L} + d' \ket{R}
\end{eqnarray*}
at each time step for which that coin is active, where {\it L} and {\it R} can be respectively thought of as the head and tail states of the coin, or equivalently as an internal chirality state of the particle. The value of the coin controls the direction in which the particle moves. When the coin shows {\it L}, the particle moves one unit to the left, when it shows {\it R}, it moves one unit to the right. Then a B-type coined QW is also defined in a similar way as we will state later. Thus the coined QW can be considered as a quantum version of a classical random walk with an additional degree of freedom called the chirality which takes values {\it left} and {\it right}. 

The amplitude of the location of the particle is defined by a 2-vector $\in {\bf C}^2$ at each location at any time $n$. The probability that the particle is at location $k$ is given by the square of the modulus of the vector at $k$. For the $j$-type coined QW ($j=A,B$), let $\ket{\Psi_{j,k} (n)}$ denote the amplitude at time $n$ at location $k$ where
\begin{eqnarray*}
\ket{\Psi_{j,k} (n)} = 
\left(
\begin{array}{cc}
\psi_{j,k} ^L (n) \\
\psi_{j,k} ^R (n)
\end{array}
\right)
\end{eqnarray*}
with the chirality being left (upper component) or right (lower component). For each $j = A$ and $B$, the dynamics of $\ket{\Psi_{j,k} (n)}$ for the $j$-type coined QW starting from the origin with an initial qubit state $\varphi = {}^t (\alpha, \beta),$ (where $\alpha, \beta \in {\bf C}$ and $|\alpha|^2 + |\beta|^2 = 1$), is presented as the following transformation:
\begin{eqnarray}
\ket{\Psi_{j,k} (n+1)} = P_j \ket{\Psi_{j,k+1} (n)} + Q_j \ket{\Psi_{j,k-1} (n)}\label{eqn:hideyoshi}
\end{eqnarray}
where
\begin{eqnarray*}
P_A= 
\left(
\begin{array}{cc}
a' & b' \\
0 & 0 
\end{array}
\right), 
\quad
Q_A=
\left(
\begin{array}{cc}
0 & 0 \\
c' & d' 
\end{array}
\right)
\end{eqnarray*}
and
\begin{eqnarray*}
P_B= 
\left(
\begin{array}{cc}
a' & 0 \\
c' & 0 
\end{array}
\right), 
\quad
Q_B=
\left(
\begin{array}{cc}
0 & b' \\
0 & d' 
\end{array}
\right)
\end{eqnarray*}
It is noted that $U=P_j+Q_j \> (j = A, B).$ The unitarity of $U$ ensures that the amplitude always defines a probability distribution for the location. From Eq. (\ref{eqn:hideyoshi}), we see that a unitary matrix of the system is described as 
\begin{eqnarray*}
\left(
\begin{array}{ccccccc}
\ddots & \vdots & \vdots & \vdots & \vdots & \vdots & \vdots \\
\ldots & O   & P_j & O   & O   & O   & \ldots \\
\ldots & Q_j & O   & P_j & O   & O   & \ldots \\
\ldots & O   & Q_j & O   & P_j & O   & \ldots \\
\ldots & O   & O   & Q_j & O   & P_j & \ldots \\
\ldots & O   & O   & O   & Q_j & O   & \ldots \\
\ldots & \vdots & \vdots & \vdots & \vdots & \vdots & \ddots
\end{array}
\right)
\qquad
\hbox{with}
\qquad
O = 
\left(
\begin{array}{cc}
0 & 0 \\
0 & 0 
\end{array}
\right)
\end{eqnarray*}
for $j = A$ and $B$. Remark that the A-type (resp. B-type) coined QW is called an A-type (resp. a G-type) quantum random walk in our previous paper \cite{KonA}. 

\section{Connection between QCA and A-type coined QW}
To begin with, we investigate a relation between the QCA and the A-type coined QW. To do so, the unitary matrix of the QCA 
\begin{eqnarray*}
\overline{U} \> = \> 
\bordermatrix{
   & \ldots & -3 & -2 & -1 & \>\>\>0 & +1 & +2 & +3 & +4 &\ldots \cr
\vdots & \ddots & \cdot  &  \cdot &  \cdot &  \cdot & \cdot &  \cdot &  \cdot & \cdot & \ldots \cr
-3 & \ldots &  b &  a &  0 & 0 &  0 &  0 &  0 & 0 & \ldots \cr
-2 & \ldots &  a &  b &  c & d &  0 &  0 &  0 & 0 & \ldots \cr
-1 & \ldots &  d &  c &  b & a &  0 &  0 &  0 & 0 & \ldots \cr
\>\>\> 0  &  \ldots & 0 &  0 &  a & b & c & d  & 0 & 0 & \ldots \cr
+1  &  \ldots & 0 &  0 &  d & c & b & a & 0 & 0 & \ldots \cr
+2  &  \ldots & 0 &  0 &  0 & 0 & a & b & c & d & \ldots \cr
+3  &  \ldots & 0 &  0 &  0 & 0 & d & c & b & a & \ldots \cr
+4  &  \ldots & 0 &  0 &  0 & 0 & 0 & 0 & a & b & \ldots \cr
\vdots & \ldots & \cdot  &  \cdot &  \cdot &  \cdot & \cdot &  \cdot &  \cdot & \cdot & \ddots 
}
\label{eqn:masaesan}
\end{eqnarray*}
is rewritten as 
\begin{eqnarray*}
\overline{U} \> = \> 
\bordermatrix{
   & \ldots & -1 & \>\>\>0 & +1 & +2 & \ldots \cr
\vdots & \ddots & \vdots & \vdots & \vdots & \vdots & \vdots \cr
-1 & \ldots & T_A & Q_A & O   & O   & \ldots \cr
\>\>\>0 & \ldots & P_A & T_A & Q_A & O   &  \ldots \cr
+1 & \ldots & O & P_A & T_A & Q_A & \ldots \cr
+2 & \ldots & O   & O   & P_A & T_A &  \ldots \cr
\vdots & \ldots & \vdots & \vdots & \vdots & \vdots & \ddots
}
\end{eqnarray*}
where 
\begin{eqnarray*}
P_A= 
\left(
\begin{array}{cc}
d & c \\
0 & 0 
\end{array}
\right),
\quad 
T_A= 
\left(
\begin{array}{cc}
b & a \\
a & b 
\end{array}
\right), 
\quad
Q_A=
\left(
\begin{array}{cc}
0 & 0 \\
c & d 
\end{array}
\right)
\end{eqnarray*}
We consider a pair $(2k-1, 2k)$ in the QCA as a site $k$ in the A-type coined QW for any $k \in {\bf Z}$. Moreover we observe that $2k-1$ (resp. $2k$) site in the QCA corresponds to right (resp. left) chirality at a site $k$ in the A-type coined QW. We call the QCA a {\it generalized A-type coined QW}. When $T_A$ is not zero matrix, the particle has non-zero amplitudes for maintaining its position during each time step. More precisely, 
\begin{eqnarray*}
\ket{\Psi_{A,k} (n)} = 
\left(
\begin{array}{cc}
\psi_{A,k} ^R (n) \\
\psi_{A,k} ^L (n)
\end{array}
\right)
\end{eqnarray*}
and
\begin{eqnarray*}
\ket{\Psi_{A,k} (n+1)} = Q_A \ket{\Psi_{A,k+1} (n)} 
+ T_A \ket{\Psi_{A,k} (n)} + P_A \ket{\Psi_{A,k-1} (n)}
\end{eqnarray*}
From the above observation, we see that ``Type V QCA $\longleftrightarrow$ generalized A-type coined QW'', where ``$X \longleftrightarrow Y$'' means that there is a one-to-one correspondence between $X$ and $Y$; that is, 
\begin{eqnarray*}
&&
\psi_{A,k} ^R (n) = \beta \eta^{(-1)} _{2k-1} (n) + \alpha \eta^{(0)} _{2k-1} (n), \quad \psi_{A,k} ^L (n) = \beta \eta^{(-1)} _{2k} (n) + \alpha \eta^{(0)} _{2k} (n)
\\
&&
x_{2k-1} ^{(0:-)} (n) = |\psi_{A,k} ^R (n)|^2, \quad x_{2k} ^{(0:-)} (n) = |\psi_{A,k} ^L (n)|^2
\end{eqnarray*}
Here we recall Type III: $|c|^2+|d|^2=1, \> c \overline{d} + \overline{c} d = 0, \> cd \not=0, \> a=b=0.$ In this case, $T_A$ becomes zero matrix. So we see that Type III is nothing but an A-type QW by interchanging $P_A$ and $Q_A,$ and the roles of left and right chiralities with $c=b'=c', \> d=a'=d'$. That is, ``Type III QCA $\longleftrightarrow$ A-type coined QW''.

We should remark that as $\tan \phi$ increases (see Eq. (\ref{eqn:nobunaga})), the relative weight of $T_A$ increases and the particle has greater probability of maintaining its position. This property also holds in a generalized B-type case introduced in the next section.

\section{Connection between QCA and B-type coined QW}
As in the case of the A-type coined QW, we study a relation between the QCA and the B-type coined QW; that is, ``Type V QCA $\longleftrightarrow$ generalized B-type coined QW''. To do this, the unitary matrix of the QCA 
\begin{eqnarray*}
\overline{U} \> = \> 
\bordermatrix{
   & \ldots & -2 & -1 & \>\>\>0 & +1 & +2 & +3 & +4 & +5  \ldots \cr
\vdots & \ddots & \cdot  &  \cdot &  \cdot &  \cdot & \cdot &  \cdot &  \cdot & \cdot & \ldots \cr
-2 & \ldots &  b &  c &  d & 0 &  0 &  0 &  0 & 0 & \ldots \cr
-1 & \ldots &  c &  b &  a & 0 &  0 &  0 &  0 & 0 & \ldots \cr
\>\>\>0 & \ldots &  0 &  a &  b & c &  d &  0 &  0 & 0 & \ldots \cr
+1 & \ldots &  0 &  d &  c & b &  a & 0 & 0 & 0 & \ldots \cr
+2 & \ldots &  0 &  0 &  0 & a & b & c & d & 0 & \ldots \cr
+3  &  \ldots & 0 &  0 &  0 & d & c & b & a & 0 & \ldots \cr
+4  &  \ldots & 0 &  0 &  0 & 0 & 0 & a & b & c & \ldots \cr
+5 & \ldots &  0 &  0 &  0 &  0 & 0 &  d &  c &  b & \ldots \cr
\vdots & \ldots & \cdot  &  \cdot &  \cdot &  \cdot & \cdot &  \cdot &  \cdot & \cdot & \ddots 
}
\label{eqn:masaesan}
\end{eqnarray*}
is rewritten as 
\begin{eqnarray*}
\overline{U} \> = \> 
\bordermatrix{
   & \ldots & -1 & \>\>\>0 & +1 & +2 & \ldots \cr
\vdots & \ddots & \vdots & \vdots & \vdots & \vdots & \vdots \cr
-1 & \ldots & T_B & P_B & O   & O   & \ldots \cr
\>\>\>0 & \ldots & Q_B & T_B & P_B & O   &  \ldots \cr
+1 & \ldots & O & Q_B & T_B & P_B & \ldots \cr
+2 & \ldots & O   & O   & Q_B & T_B &  \ldots \cr
\vdots & \ldots & \vdots & \vdots & \vdots & \vdots & \ddots
}
\end{eqnarray*}
where 
\begin{eqnarray*}
P_B= 
\left(
\begin{array}{cc}
d & 0 \\
a & 0 
\end{array}
\right),
\quad 
T_B= 
\left(
\begin{array}{cc}
b & c \\
c & b 
\end{array}
\right), 
\quad
Q_B=
\left(
\begin{array}{cc}
0 & a \\
0 & d 
\end{array}
\right)
\end{eqnarray*}
We consider a pair $(2k, 2k+1)$ in the QCA as a site $k$ in the B-type coined QW for any $k \in {\bf Z}$. Moreover we observe that $2k$ (resp. $2k+1$) site in the QCA corresponds to left (resp. right) chirality at site $k$ in the B-type coined QW. We call the QCA a {\it generalized B-type coined QW}. When $T_B$ is not zero matrix, the particle has non-zero amplitudes for maintaining its position during each time step. As in the case of the A-type QW, it is shown that ``Type V QCA $\longleftrightarrow$ generalized B-type coined QW'', that is, 
\begin{eqnarray*}
&&
\psi_{B,k} ^L (n) = \alpha \eta^{(0)} _{2k} (n) + \beta \eta^{(1)} _{2k} (n), \quad \psi_{B,k} ^R (n) = \alpha \eta^{(0)} _{2k+1} (n) + \beta \eta^{(1)} _{2k+1} (n)
\\
&&
x_{2k} ^{(0:+)} (n) = |\psi_{B,k} ^L (n)|^2, \quad x_{2k+1} ^{(0:+)} (n) = |\psi_{B,k} ^R (n)|^2
\end{eqnarray*}
We think of Type IV: $|a|^2+|d|^2=1, \> a \overline{d} + \overline{a} d = 0, \> ad \not=0, \> b=c=0.$ In this case, $T_B$ is zero matrix. So Type IV becomes a B-type coined QW with $d=a'=d', \> a=b'=c'$; that is, ``Type IV QCA $\longleftrightarrow$ B-type coined QW''.

Meyer \cite{MeyerA,MeyerB,MeyerC} has investigated the B-type coined QWs, which was called a {\it quantum lattice gas automaton}. His case (for example, Eq. (24) in his paper \cite{MeyerA}) can be obtained by $\delta \to 3 \pi/2, \> \phi \to \rho,$ and $\theta \to \pi/2 + \theta$ in Eq. (\ref{eqn:nobunaga}).

\section{Connection between Type V QCA and two-step coined QW}
In the previous two sections, we have clarified the following relations: ``Type III QCA $\longleftrightarrow$ A-type coined QW'', ``Type IV QCA $\longleftrightarrow$ B-type coined QW'', moreover ``Type V QCA $\longleftrightarrow$ generalized A-type coined QW $\longleftrightarrow$ generalized B-type coined QW''. This section gives a relation between Type V QCA and two-step coined QW. The meaning of the ``two-step'' is that we identify the one-step transition of Type V QCA with the two-step transition of two-step coined QW. 

First ``Type V QCA $\longleftrightarrow$ two-step A-type coined QW'' is given. Next ``Type V QCA $\longleftrightarrow$ two-step B-type coined QW'' is also presented. Combining them all, we finally obtain the following relations:
\par\noindent
``Type V QCA $\longleftrightarrow$ generalized A-type coined QW $\longleftrightarrow$ two-step A-type coined QW''
\par\noindent
``Type V QCA $\longleftrightarrow$ generalized B-type coined QW $\longleftrightarrow$ two-step B-type coined QW''

Now we present ``generalized A-type coined QW $\longleftrightarrow$ two-step A-type coined QW'' in the following way. A direct computation implies that a generalized A-type coined QW with
\begin{eqnarray*}
P_A= 
\left(
\begin{array}{cc}
d & c \\
0 & 0 
\end{array}
\right),
\quad 
T_A= 
\left(
\begin{array}{cc}
b & a \\
a & b 
\end{array}
\right), 
\quad
Q_A=
\left(
\begin{array}{cc}
0 & 0 \\
c & d 
\end{array}
\right)
\end{eqnarray*}
is equivalent to a two-step A-type coined QW with 
\begin{eqnarray*}
P_A (1) = 
\left(
\begin{array}{cc}
i \cos \phi \> e^{i \theta_2}  & \sin \phi \> e^{i \theta_2}  \\
0 & 0 
\end{array}
\right),
\quad 
P_A (2) = e^{i \delta}
\left(
\begin{array}{cc}
\sin \theta \> e^{-i \theta_2}  & -i \cos \theta \> e^{i \theta_1} \\
0 & 0 
\end{array}
\right) 
\end{eqnarray*}
and
\begin{eqnarray*}
Q_A (1) = 
\left(
\begin{array}{cc}
0 & 0  \\
\sin \phi \> e^{i \theta_1} & i \cos \phi \> e^{i \theta_1} 
\end{array}
\right),
\quad 
Q_A (2) = e^{i \delta}
\left(
\begin{array}{cc}
0 & 0 \\
-i \cos \theta \> e^{-i \theta_2} & \sin \theta \> e^{-i \theta_1}  
\end{array}
\right) 
\end{eqnarray*}
for any $\theta_1, \theta_2 \in [0, 2 \pi)$ where 
\begin{eqnarray*}
P_A = P_A (2) P_A (1), \quad Q_A = Q_A (2) Q_A (1), 
\quad T_A = P_A (2) Q_A (1) + Q_A (2) P_A (1)
\end{eqnarray*}
Note that $(a,b,c,d)$ has the form given in Eq. (\ref{eqn:nobunaga}) and $U (n) \equiv P_A (n) + Q_A (n)$ is unitary for $n=1,2.$

In a similar fashion, we show that ``generalized B-type coined QW $\longleftrightarrow$ two-step B-type coined QW''; that is, a generalized B-type coined QW with 
\begin{eqnarray*}
P_B= 
\left(
\begin{array}{cc}
d & 0 \\
a & 0 
\end{array}
\right),
\quad 
T_B= 
\left(
\begin{array}{cc}
b & c \\
c & b 
\end{array}
\right), 
\quad
Q_B=
\left(
\begin{array}{cc}
0 & a \\
0 & d 
\end{array}
\right)
\end{eqnarray*}
corresponds to a two-step B-type coined QW with 
\begin{eqnarray*}
P_B (1) = 
\left(
\begin{array}{cc}
i \cos \phi \> e^{i \theta_2}  & 0  \\
\sin \phi \> e^{i \theta_1}    & 0 
\end{array}
\right),
\> 
P_B (2) = e^{i \delta}
\left(
\begin{array}{cc}
\sin \theta \> e^{-i \theta_2}     & 0 \\
-i \cos \theta \> e^{-i \theta_2}  & 0 
\end{array}
\right) 
\end{eqnarray*}
and
\begin{eqnarray*}
Q_B (1) = 
\left(
\begin{array}{cc}
0 & \sin \phi \> e^{i \theta_2}  \\
0 & i \cos \phi \> e^{i \theta_1} 
\end{array}
\right),
\>
Q_B (2) = e^{i \delta}
\left(
\begin{array}{cc}
0 & -i \cos \theta \> e^{i \theta_1} \\
0 & \sin \theta \> e^{-i \theta_1}  
\end{array}
\right)
\end{eqnarray*}
for any $\theta_1, \theta_2 \in [0, 2 \pi)$ where 
\begin{eqnarray*}
P_B = P_B (2) P_B (1), \quad Q_B = Q_B (2) Q_B (1), 
\quad T_B = P_B (2) Q_B (1) + Q_B (2) P_B (1)
\end{eqnarray*}
Remark that $P_B (n) + Q_B (n)= P_A (n) + Q_A (n)$ for $n=1,2.$

From now on we discuss the case given by Patel {\it et al.} \cite{Patel}. In their notation, we take general $2 \times 2$ blocks of $U_e$ and $U_o$ as 
\begin{eqnarray*}
U_e =
\left(
\begin{array}{cc}
\cos \phi_1 & i \sin \phi_1   \\
i \sin \phi_1  & \cos \phi_1  
\end{array}
\right), 
\qquad
U_o =
\left(
\begin{array}{cc}
\cos \phi_2 & i \sin \phi_2   \\
i \sin \phi_2  & \cos \phi_2  
\end{array}
\right)
\end{eqnarray*}
Their special case is $\phi_1=\phi_2=\pi/4,$ i.e., 
\begin{eqnarray}
U_e = U_o =
{1 \over \sqrt{2}}
\left(
\begin{array}{cc}
1 & i   \\
i & 1  
\end{array}
\right)
\label{eqn:ai}
\end{eqnarray}
By using $U_e$ and $U_o$, the following matrices are defined:
\begin{eqnarray*}
\overline{U}_e \> = \> 
\bordermatrix{
   & \ldots & -2 & -1 & \>\>\>0 & +1 & +2 & +3 & \ldots \cr
\vdots & \ddots & \cdot &  \cdot &  \cdot & \cdot & \cdot & \cdot & \ldots \cr
-2 & \ldots & \cos \phi_1 & i \sin \phi_1 & 0 & 0 & 0 & 0 & \ldots \cr
-1 & \ldots & i \sin \phi_1 & \cos \phi_1 & 0 & 0 & 0 & 0 &  \ldots \cr
\>\>\> 0  &  \ldots & 0 &  0 &  \cos \phi_1 & i \sin \phi_1 & 0 & 0 & \ldots 
\cr 
+1  &  \ldots & 0 &  0 &  i \sin \phi_1 & \cos \phi_1 & 0 & 0 & \ldots \cr
+2  &  \ldots & 0 &  0 &  0 & 0 & \cos \phi_1 & i \sin \phi_1 & \ldots \cr
+3  &  \ldots & 0 &  0 &  0 & 0 & i \sin \phi_1 &  \cos \phi_1 & \ldots \cr
\vdots & \ldots & \cdot  &  \cdot &  \cdot &  \cdot & \cdot & \cdot & \ddots 
}
\label{eqn:atsui}
\end{eqnarray*}
and
\begin{eqnarray*}
\overline{U}_o \> = \> 
\bordermatrix{
   & \ldots & -2 & -1 & \>\>\>0 & +1 & +2 & +3 & \ldots \cr
\vdots & \ddots & \cdot &  \cdot &  \cdot & \cdot & \cdot & \cdot & \ldots \cr
-2 & \ldots & \cos \phi_2 & 0 & 0 & 0 & 0 & 0 & \ldots \cr
-1 & \ldots & 0 & \cos \phi_2 & i \sin \phi_2 & 0 & 0 & 0 &  \ldots \cr
\>\>\> 0  &  \ldots & 0 &  i \sin \phi_2 &  \cos \phi_2 & 0 & 0 & 0 & \ldots 
\cr 
+1  &  \ldots & 0 &  0 &  0 & \cos \phi_2 & i \sin \phi_2 & 0 & \ldots \cr
+2  &  \ldots & 0 &  0 &  0 & i \sin \phi_2 & \cos \phi_2 & 0 & \ldots \cr
+3  &  \ldots & 0 &  0 &  0 & 0 & 0 &  \cos \phi_2 & \ldots \cr
\vdots & \ldots & \cdot  &  \cdot &  \cdot &  \cdot & \cdot & \cdot & \ddots 
}
\label{eqn:atsuiyo}
\end{eqnarray*}
Noting that $\overline{U}=\overline{U}_e \overline{U}_o$, we have 
\begin{eqnarray}
(a,b,c,d) = 
(i \cos \phi_1 \sin \phi_2, \cos \phi_1 \cos \phi_2, 
 i \sin \phi_1 \cos \phi_2, - \sin \phi_1 \sin \phi_2)
\label{eqn:seika}
\end{eqnarray}
Therefore, by choosing $\theta = \phi_1, \phi = \pi/2 - \phi_2,$ and $\delta = \pi/2$ in  Eq. (\ref{eqn:nobunaga}), we obtain Eq. (\ref{eqn:seika}).

Furthermore, we see that if $\theta + \phi = \pi/2, 3\pi/2, \> \theta_1 = \theta_2,$ and $\delta = 2 \theta_1 + \pi/2,$ then $U (1) = U (2).$ To get the case of Patel {\it et al.}, we take $\theta = \phi = \pi/4, \> \theta_1=\theta_2=0,$ and $\delta =\pi/2$, so 
\begin{eqnarray}
U (1) = U(2) = {1 \over \sqrt{2}}
\left(
\begin{array}{cc}
i & 1  \\
1 & i 
\end{array}
\right) 
\label{eqn:aiyo}
\end{eqnarray}
Remark that $U_e = U_o$ is not equal to $U(1)=U(2)$ in their case (see Eqs. (\ref{eqn:ai}) and (\ref{eqn:aiyo})). To obtain their asymptotic result, we define their walk at time $n$ with the initial qubit state $\varphi = {}^t (1/\sqrt{2}, 1/\sqrt{2})$ by $X_n ^{\varphi}.$ Note that if $\varphi = {}^t (\alpha , \beta)$ satisfies $\alpha \overline{\beta} = \overline{\alpha} \beta,$ then the distribution is symmetric at any time, (see Theorem 4 in \cite{KonB}). Then, our limit theorem \cite{KonA,KonB,KonC} implies that 
\begin{eqnarray*}
P(a \le X^{\varphi} _{n}/n \le b)  \to  \int^b _a  {4 \over \pi (4 - x^2) \sqrt{4 - 2 x^2}} dx \qquad (n \to \infty) 
\end{eqnarray*}  
for $ -\sqrt{2} \le a < b \le \sqrt{2}$. It should be noted that their case can be considered as a two-step coined QW with $U(1)=U(2)$, so we make a change of variables; $x \to x/2$ in our original papers \cite{KonA,KonB,KonC}. The above limit density function corresponds to Eq. (34) at time $t=1$ in their paper \cite{Patel}. Thus, their asymptotic result can be easily derived from the connection between the QCA and the two-step coined QW that is given in this section. Moreover, it would be shown that a similar convergence theorem holds for any general model with $U(1)=U(2)$ as in the above case.
\par
Finally we should remark that in more general setting, Severini \cite{Seve} has studied combinatorial properties of the digraphs of unitary matrices to clarify the following question: On which digraphs can QWs be defined ?



\begin{small}

\end{small}

\end{document}